\documentclass[aps,prb,twocolumn,amsmath,amssymb,showpacs]{revtex4-1}
\usepackage{graphicx}
\usepackage{graphics}
\usepackage{amsfonts,epsfig}
\usepackage{float}
\usepackage{color}
\allowdisplaybreaks

\begin{document}

\title{Exact magnetic field control of nitrogen-vacancy center spin for realizing fast quantum logic gates}

\author{Wen-Qi Fang and Bang-Gui Liu}
\email{bgliu@iphy.edu.cn} \affiliation{Beijing National Laboratory
for Condensed Matter Physics, Institute of Physics, Chinese Academy
of Sciences, Beijing 100190, China.}

\date{\today}

\begin{abstract}
The negatively charged nitrogen-vacancy (NV) center spin in diamond can be used to realize quantum computation and to sense magnetic fields. Its spin triplet consists of three levels labeled with its spin z-components of +1, 0, and -1. Without external field, the +1 and -1 states are degenerate and higher than the 0 state due to the zero-field splitting. By taking the symmetrical and anti-symmetrical superpositions of the +1 and -1 states as our qubit basis, we obtain exact evolution operator of the NV center spin under time-dependent magnetic field by mapping the three-level system on time-dependent quantum two-level systems with exact analytical solutions. With our exact evolution operator of the NV center spin including three levels, we show that arbitrary qubits can be prepared from the starting 0 state and arbitrary rapid quantum logic gates of these qubits can be realized with magnetic fields. In addition, it is made clear that the typical quantum logic gates can be accomplished within a few nanoseconds and the fidelity can be very high because only magnetic field strength needs to be controlled in this approach. These results should be useful to realizing quantum computing with the NV center spin systems in diamond and exploring other effects and applications.
\end{abstract}


\maketitle

\section{Introduction}

The negatively charged nitrogen-vacancy (NV) center in diamond has been intensively investigated because it can be used for realizing quantum computation and sensing weak magnetic field, electric field, strain etc.\cite{dobro,schir,doher,mans,info}
As quantum technology evolves,
one can manipulate the NV center spin with electromagnetic field
\cite{fuchs,lond,mayer}, optical field \cite{tama,bass,jacq}, and
stress field \cite{macq,teis}. Its spin ground state is a
triplet ($S=1$) with 0 and $\pm 1$ as the spin z exponents, and its zero-field splitting makes the $|0\rangle$ state the lowest, leaving the $|+1\rangle$ and $|-1\rangle$ degenerate. One can apply a longitudinal magnetic field to
split the $|+1\rangle$ and $|-1\rangle$ states, adding a Zeeman term to
the spin Hamiltonian. Usually, it is preferable to work with a two level system
representing a qubit by using the $|0\rangle$ state and one of  the $|+1\rangle$ and $|-1\rangle$ states, and then control it with microwave pulse or radio-frequency wave \cite{fuchs}. In such approaches, however, one usually use only weak driving under the rotating wave approximation, or else the third state cannot be neglected. Additionally, one needs accurate timing to track the relative phase between the $|0\rangle$ and $|+1\rangle$ (or $|+1\rangle$) states due to the zero-field splitting effect.
Recently, the NV center spin was used to experimentally realize geometric quantum gates with high fidelity\cite{arroyo,czu}. In this approach, the $|+1\rangle$ and $|-1\rangle$ states are used to encode the qubit. The latest report shows that the $|+1\rangle$ and $|-1\rangle$ states can be manipulated with internal strain field\cite{mech}. Nevertheless, the geometric quantum gates need half microsecond to be accomplished and the mechanical quantum gates require longer time. Therefore, much more rapid quantum logic gates of such qubits are highly desirable.

In this paper, we use  $|\pm \rangle=(|+1\rangle \pm|-1\rangle)/\sqrt{2}$ of the NV center spin as the qubit basis and then show how to prepare arbitrary qubits from the starting $|0\rangle$ state and how to construct rapid quantum logic gates for the qubits. On the basis of powerful exact analytical results of a quantum two-level system under a time-dependent magnetic field\cite{barnes,barnes1}, we construct exact evolution operator for the NV center spin (including three levels) under time-dependent magnetic fields. Then, we use these exact evolution operators to prepare arbitrary qubit states with the basis $|\pm \rangle$ from the starting state $|0\rangle$ and realize arbitrary quantum logic gates.
All the typical quantum logic gates can be completed within a few nanoseconds. The fidelity can be made very high because one needs to control magnetic field strength only. More detailed results will be presented in the following.

The rest of the paper is organized as follows. In Set. \textrm{II}, we define the Hamiltonian and elucidate the new spin basis. In Sec. \textrm{III}, we construct exact evolution operators for the NV center spin under time-dependent magnetic fields. In Sec. \textrm{IV}, we show how to use the exact evolution operators to prepare arbitrary qubit states with the basis $|\pm \rangle$ from the starting state $|0\rangle$. In Sec. \textrm{V}, we construct arbitrary quantum logic gates for the NV center spin qubits by using the exact evolution operators. Finally, we make some necessary discussions and give our conclusion in Sec. \textrm{VI}.

\section{New spin basis}

The Hamiltonian of NV center spin $\vec{S}$, in the presence of time
dependent magnetic field $\vec{B}(t)$=$(B_x(t),B_y(t),B_z(t))$, can be
written as
\begin{equation}\label{eq1}
H=DS_{z}^{2}+\gamma \vec{S}\cdot\vec{B}(t),
\end{equation}
where $\hbar=1$ is used, $D=2.87$GHz is the zero-field splitting,
and $\gamma=2.8$MHz/G is the electron gyromagnetic ratio. Accordingly, the
Schr\"odinger equation for time-evolution operator $U$ is given by
$i\frac{d}{dt}U=HU$.

In the $S_z$ representation, the matrix form of the Hamiltonian
(\ref{eq1}) can be written as:
\begin{equation}\label{eq2}
H=\left(\begin{array}{ccc}
D+\gamma B_z & \frac{\gamma  (B_x-i B_y)}{\sqrt{2}} & 0 \\
\frac{\gamma (B_x+i B_y)}{\sqrt{2}} & 0 & \frac{\gamma (B_x-i B_y)}{\sqrt{2}} \\
0 & \frac{\gamma  (B_x+i B_y)}{\sqrt{2}} & D-\gamma B_z \\
\end{array}\right).
\end{equation}
Introducing the unitary transform
\begin{equation}\label{eq3}
S_{1}=\left(
\begin{array}{ccc}
\frac{1}{\sqrt{2}}& 0 & \frac{1}{\sqrt{2}} \\
0 & 1 & 0 \\
\frac{1}{\sqrt{2}} & 0 & -\frac{1}{\sqrt{2}}
\end{array}
\right),
\end{equation}
we can transform the Hamiltonian (\ref{eq2}) into
\begin{equation}\label{eq4}
H^{'}=\left(\begin{array}{ccc}
D & \gamma B_x & \gamma B_z \\
\gamma B_x & 0 & i\gamma B_y \\
\gamma B_z & -i\gamma B_y & D \\
\end{array}\right).
\end{equation}
It means that the basis is changed from ($|+1\rangle$,$|0\rangle$,$|-$1$\rangle$)
to ($|+\rangle$,$|0\rangle$,$|-\rangle$), where $|+\rangle$=$\frac{1}{\sqrt 2}$($|+1\rangle
+|-1\rangle$) and $|-\rangle$=$\frac{1}{\sqrt2}$($|+1\rangle-|-1\rangle$). When the
magnetic field is turned off, the $D$ term will produce the same overall phase
for both $|+\rangle$ and $|-\rangle$. Therefore, there will be no relative
phase between $|+\rangle$ and $|-\rangle$, and $|+\rangle$ and $|-\rangle$ can be
used to make a stable qubit basis.

\section{exact evolution operators}

With a special magnetic field $\vec{B}(t)$=$(\alpha B_1,\beta B_1,0)$, the Hamiltonian in Eq. (\ref{eq4}) becomes
\begin{equation}\label{eq5}
H_{o}=\left(\begin{array}{ccc}
D & \alpha\gamma B_1 & 0 \\
\alpha\gamma B_1 & 0 & i\beta\gamma B_1 \\
0 & -i\beta\gamma B_1 & D \\
\end{array}\right),
\end{equation}
where $\alpha$ and $\beta$ are adjustable real parameters satisfying $\alpha^{2}+\beta^{2}$=$1$. Because of this condition, $\alpha$ and $\beta$ can be parameterized as $\alpha=\cos \theta$ and $\beta=\sin \theta$, where $-\pi \le \theta \le \pi$.
Introducing another unitary transform
\begin{equation}\label{eq6}
S_{2}=\left(\begin{array}{ccc}
\cos \theta  & 0 & i \sin\theta \\
 0 & 1 & 0 \\
i \sin\theta & 0 & \cos\theta \\
\end{array}\right)
\end{equation}
we can transform $H_o$ into a block-diagonal matrix $H_t=S_2H_oS_2^{\dag}$,
\begin{equation}\label{eq7}
H_{t}=\left(\begin{array}{ccc}
D & J(t) & 0 \\
J(t) & 0 & 0 \\
0 & 0 & D \\
\end{array}\right)
\end{equation}
where $J(t)$=$\gamma B_1(t)$.
Because $H_t$ consists of a 2$\times$2 block
\begin{equation}\label{eq72}
H_{t2}=\left(\begin{array}{cc}
D & J(t)  \\
J(t) & 0  \\
\end{array}\right)
\end{equation}
and a 1$\times$1 block $H_{t1}=D$,
we can focus on $H_{t2}$ in Eq. (\ref{eq72}).

Introducing a 2$\times$2 unitary transform
\begin{equation}
T_{h}= \frac{1}{{\sqrt 2
}}\left( {\begin{array}{*{20}{c}}
   1 & 1  \\
   1 & { - 1}  \\
\end{array}} \right),
\end{equation}
we can change $H_{t2}$ into $H_{h2}=H_2+D/2$, with $H_2$ given by
\begin{equation}\label{eqH2}
H_{2}=\left(
{\begin{array}{*{20}{c}}
   { J(t)} & {\frac{D}{2}}  \\
   {\frac{D}{2}} & { -  J(t)}  \\
\end{array}} \right).
\end{equation}
The constant term $\frac{D}{2}$ in $H_{h2}$ contributes only an overall phase to the time
evolution operator. For the time-dependent two-level Hamiltonian $H_{2}$, there are many exact solutions such as those\cite{add1,add2,add3,barnes,barnes1}. Here, we choose a powerful method to construct its exact time evolution operator\cite{barnes,barnes1}.

The Schr\"odinger equation of $H_{2}$ can be exactly solved by the evolution operator\cite{barnes,barnes1},
\begin{equation}\label{eq8}
U_2=\left(\begin{array}{ccc}
u_{11} & -u_{21}^{\ast } \\
u_{21} & u_{11}^{\ast }
\end{array}\right), |u_{11}|^2+|u_{21}|^2=1,
\end{equation}
and the matrix elements $u_{11}$ and $u_{21}$ and the quantity $J(t)$ can be expressed as\cite{barnes,barnes1}
\begin{equation}\label{eq9}\left\{
\begin{array}{l}
{u_{11}}(t)=\cos (\chi (t)){e^{i{\xi_-}(t)}} \\
{u_{21}}(t) = i\eta \sin (\chi (t)){e^{i{\xi_+}(t)}} \\
J(t) = \frac{\ddot{\mathop {\chi}}(t)}{{\sqrt {{D^2} -
4\dot{\mathop {\chi }}(t)^2 } }} - \frac{1}{2}\sqrt {{D^2} -
4\dot{\mathop {\chi }}(t)^2 } \cot (2\chi (t))
\end{array}\right.,
\end{equation}
where $\xi_{\pm}$ is defined as
\begin{equation}\label{eq10}
\xi_{\pm}=\int_{0}^{t}dt^{^{\prime
}}\frac{1}{2}\sqrt{D^{2}-4\dot{\chi}^{2}}\csc (2\chi)\pm
\frac{1}{2}\arcsin (\frac{2\dot{\chi}}{D})\pm \eta \frac{\pi }{4}.
\end{equation}
Here, $\eta$ can take either $+1$ or $-1$, and $\chi$ must satisfies three conditions:
$|{\dot{\mathop{\chi}}(t)}|\le\frac{D}{2}$, $\chi(0)=0$, and
$\dot{\mathop{\chi}}(0)=-\eta \frac{D}{2}$. By choosing suitable $\chi(t)$, we can exactly construct $J(t)$ in $H_2$ and the evolution operator $U_2$ in this way.

After adding the phase factor due to the $D/2$ term and making the inverse unitary
transformation with $T_h^{\dagger}$, we obtain the evolution operator for $H_{t2}$,
\begin{equation}\label{Ut2}
U_{t2}\left( t\right) =\left(
\begin{array}{cc}
\bar{u}_{11} & -\bar{u}_{21}^{\ast }  \\
\bar{u}_{21} & \bar{u}_{11}^{\ast }  \\
\end{array}%
\right) e^{-i\frac{D}{2}t}
\end{equation}
where the matrix elements $\bar{u}_{11}$ and $\bar{u}_{21}$ are expressed as
\begin{displaymath}\left\{
\begin{array}{c}
{\bar{u}_{11}}(t)=\cos(\chi(t))\cos({\xi_-}(t))+ i\eta\cos({\xi_+}(t))\sin(\chi(t))\\
{\bar{u}_{21}}(t)=\eta\sin(\chi(t))\sin({\xi_+}(t))+i\cos (\chi(t))\sin ({\xi_-}(t))\\
\end{array}\right.
\end{displaymath}

Consequently, the whole time evolution
operator of the Hamiltonian $H_{t}$ can be written as:
\begin{equation}\label{+eq11}
U_{t}\left( t\right) =\left(
\begin{array}{ccc}
\bar{u}_{11} & -\bar{u}_{21}^{\ast } & 0 \\
\bar{u}_{21} & \bar{u}_{11}^{\ast } & 0 \\
0 & 0 & e^{-i\frac{D}{2}t}%
\end{array}%
\right) e^{-i\frac{D}{2}t}
\end{equation}
After making the inverse transform $S_2^{\dagger}$, we can get the time evolution operator of the starting Hamiltonian $H_{o}$ in the new basis of $|+\rangle$ and $|-\rangle$:
\begin{equation}\label{eq12}
\begin{split}
&U_{o}(\theta,t)=d(t)\times \\
&\left(
\begin{array}{ccc}
\bar{u}_{11}\alpha^2+d(t)\beta^2 & -\bar{u}_{21}^{*}\alpha &
-i\left(d(t)-\bar{u}_{11}\right)\alpha\beta \\
\bar{u}_{21}\alpha & \bar{u}_{11}^{*} & i\bar{u}_{21}\beta \\
i\left(d(t)-\bar{u}_{11}\right)\alpha\beta & i \bar{u}_{21}^{*}\beta
& d(t)\alpha^2+\bar{u}_{11}\beta^2 \\
\end{array}
\right),
\end{split}
\end{equation}
where $\alpha=\cos \theta$, $\beta=\sin \theta$, and $d(t)=e^{-i\frac{D}{2}t}$.

From above equations (\ref{eq9}) and (\ref{eq10}), we can see that if
the time-dependent function $\chi(t)$ is specified, the evolution operator (\ref{eq12})
will be determined immediately. Using this time evolution operators,
we can control a single NV center spin exactly and efficiently,
and thus we can initialize and gate arbitrary
qubits in the stable basis of $|+\rangle$ and $|-\rangle$.

\section{Exact initializing of arbitrary qubits}

Experimentally, the NV center spin can be easily prepared in state
$|0\rangle$. We try to realize state transfer between $|0\rangle$ and $|\pm\rangle$.
With the time evolution operator $U_{o}$ applied, the state $|0\rangle$ will become
\begin{equation}\label{eq13}
U_{o}(\theta,t)|0\rangle=d(t)\left(
\begin{array}{ccc}
-\cos\theta(\eta\sin\chi\sin\xi_{+}-i\cos\chi\sin\xi_{-}) \\
\cos\chi\cos\xi_{-}-i\eta\sin\chi\cos\xi_{+}  \\
i\sin\theta(\eta\sin\chi\sin\xi_{+}-i\cos\chi\sin\xi_{-}) \\
\end{array}
\right).
\end{equation}
We require that the function $\chi(t)$ is given by
\begin{equation}\label{eq14}
\chi\left(t\right)=\lambda t-\frac{{2\kappa{\lambda^3}{t^3}}}{3}+\frac{{\kappa{\lambda^3}{t^4}}}{T_f}-\frac{{2\kappa{\lambda^3}{t^5}}}{{5{T_f^2}}},
\end{equation}
where $\lambda$ is defined as $\frac{D}{2}$, $\kappa$ is an adjustable parameter,  and $T_f$
describes the time duration. Using $\eta$=$-1$, we have $\xi_{+}(T_f)$=$\xi_{-}(T_f)$
according to equation (\ref{eq10}). Because the target state doesn't contain state $|0\rangle$, we need to set $\bar{u}_{11}(T_f)=0$, {\it i.e.} $\cos(\xi_{+}(T_f))=\cos(\xi_{-}(T_f))=0$. Then the quantity $\chi(T_f)$ contributes an overall phase in the state $U_o(\theta,T_f)|0\rangle$ in Eq. (\ref{eq13}).
In order to achieve a minimal time value $T_f$ and a finite field pulse in the time interval $t\in(0, T_f)$,
we need two conditions: $0<\chi(T_f)\leq\frac{\pi}{2}$ and
\begin{equation}\label{eq15}
\int_{0}^{T_f}dt\frac{1}{2}\sqrt{D^{2}-4\dot{\chi}^{2}}\csc (2\chi)=\frac{\pi}{2}.
\end{equation}
Once we set $t=T_f$ and choose a value for $\chi(T_f)$,
$\kappa$ can be formally solved by using equation (\ref{eq14}), reading $\kappa=15(\lambda T_f-\chi(T_f))/(\lambda T_f)^3$.
The time duration $T_f$ can be solved from equation (\ref{eq15}), and then $\kappa$ can be calculated immediately.
In the following, we shall show how to initialize three typical qubits from the spin state $|0\rangle$.

{\it Initializing the basis states} $|\pm\rangle$.
Choosing $\alpha=0$ (or $\beta=0$) in Eq. (5), we can easily get the target state $U_o(\frac{\pi}{2},T_f)|0\rangle=|-\rangle$ (or $U_o(0,T_f)|0\rangle=|+\rangle$) with an overall phase. In this way, we get $|\pm\rangle$ from $|0\rangle$.

{\it Initializing a superposed state} $\cos\theta_1|+\rangle +i\sin\theta_1|-\rangle$.
In this cases, we can assume $0\leq\theta_{1} \leq \pi$ without losing any effective information.
Choosing $\alpha$ and $\beta$ in Eq. (5) to satisfy the equality $\arctan(\beta/\alpha)=\theta \ge 0$, we can let $\theta = \pi-\theta_1$ in Eq. (17) and thus obtain the final state
\begin{equation}\label{eq16}
U_{o}(\pi-\theta_1,T_f)|0\rangle=-i\left(
\begin{array}{ccc}
\cos\theta_1\\
0  \\
i\sin\theta_1 \\
\end{array}
\right)e^{-i\chi(T_f)}e^{-i\frac{D}{2}T_f}.
\end{equation}
Neglecting the overall factor, we obtain $\cos\theta_1|+\rangle +i\sin\theta_1|-\rangle$.

{\it Initializing arbitrary state} $\cos\theta_{1}|+\rangle +e^{i\varphi}\sin\theta_{1}|-\rangle$.
We assume that $0\leq\varphi\leq\pi$. State $\cos\theta_1|+\rangle +i\sin\theta_1|-\rangle$ can be obtained
from $|0\rangle$ in the same way as above. The next step is to realize the phase factor $e^{i(\varphi-\frac{\pi}{2})}$ for $|-\rangle$.
If $B_1(t)$ in Eq. (\ref{eq5}) is a time-independent magnetic field $B_0$, the method presented in Sec. \textrm{III} is not applicable, but it is not difficult to derive the evolution operator:
\begin{equation}\label{eq17}
\begin{split}
&P(\tau,\alpha_0,\beta_0)=d(\tau)\times \\
&\left(
\begin{array}{ccc}
\beta_0 ^2 d(\tau)+(-1)^n\alpha_0 ^2& 0 & i \alpha_0\beta_0\left((-1)^n- d(\tau)\right) \\
0 & (-1)^n & 0 \\
-i\alpha_0\beta_0\left((-1)^n-d(\tau)\right) & 0 & \alpha_0^2 d(\tau)+(-1)^n\beta_0^2 \\
\end{array}
\right),
\end{split}
\end{equation}
where $n$ is a non-negative integer. We assume $\tau$=$\tau_f$ when the evolution ends. Here, $\tau_f$ must satisfy the equality: $\frac{D}{2}\tau_f\sqrt{1+4 (\frac{\gamma B_{0}}{D})^2}$=$n\pi$.
If $n=0$, we must have $\tau_f=0$. Consequently, it means $\varphi-\frac{\pi}{2}=0$, and the operator $P(\tau_f,\alpha_0,\beta_0)$ becomes identity operator, or trivial operator. For any nontrivial evolution operator, we have $\tau_f>0$ and $n=1$, Because of $\varphi-\frac{\pi}{2}\in[-\frac{\pi}{2},\frac{\pi}{2}]$, we need nontrivial evolution operators for $|\varphi-\frac{\pi}{2}|>0$.
It is easily seen that only if $\alpha_0\beta_0=0$ in (21), can arbitrary relative phases be realized by
the operator $P(\tau_f,\alpha_0,\beta_0)$. This means that the magnetic field should be applied in either
$x$-axis or $y$-axis.
For $\varphi\in[0,\frac{\pi}{2})$, we need to set $\alpha_0=0$ and then obtain $\tau_f=\frac{2}{D}(\varphi+\frac{\pi}{2})$ and $B_0=\frac{D}{2\gamma}\sqrt{(\frac{\pi}{\varphi+\frac{\pi}{2}})^2-1}$; and for $\varphi\in(\frac{\pi}{2},\pi]$, we need to set $\beta_0=0$ and then obtain $\tau_f=\frac{2}{D}(\frac{3\pi}{2}-\varphi)$ and $B_0=\frac{D}{2\gamma}\sqrt{(\frac{\pi}{\frac{3\pi}{2}-\varphi})^2-1}$. We can parameterize $\alpha_0=\cos \theta_0$ and $\beta_0=\sin \theta_0$.
In this way, time duration $\tau_f$ will be only few nanoseconds and magnetic field $B_0$ will be reasonable at the same time. Because $\theta_0$ ($\alpha_0$ and $\beta_0$), $\tau_f$, and $B_0$ are determined by $\varphi$, we can use $P_f(\varphi-\frac{\pi}{2})$ to denote the evolution operator.

Therefore, the whole procedure, achieved in two steps, can be represented as the evolution operator
\begin{equation}\label{ui}
U_I(\varphi,\theta_1)=P_f(\varphi-\frac{\pi}{2})U_{o}(\pi-\theta_1,T_f)
\end{equation}
For practical application, it is useful to adjust the value $\chi(T_f)$ to connect the two magnetic fields $B_1(t)$
and $B_0$ at the time $T_f$ along either $x$-axis or $y$-axis. We show in Fig. 1 that magnetic field $B_1(T_f)$ can vary from
$450G$ to $740000G$ when $\chi(T_f)$ changes within $(0, \frac{\pi}{2}]$. Because of the large domain of $B_1(T_f)$, we can likely realize continuous connection of magnetic field in either $x$-axis or $y$-axis.

\begin{figure}
\begin{center}
\includegraphics[width=0.8\columnwidth]{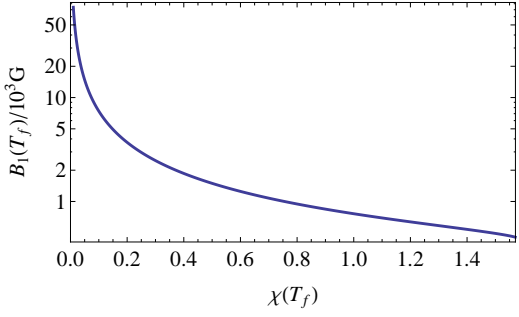}
\caption{\label{fig:qtest1} The final magnetic field $B_1(T_f)$ in unit of $10^3$G depending on parameter $\chi(T_f)$.}
\end{center}
\end{figure}

\section{realization of quantum logic gates}

In last section, we show how to initiate the basis states $|\pm\rangle$ and their superposed states from the state $|0\rangle$. Our arbitrary qubits are made from the basis states $|\pm\rangle$. In the quantum circuit model of
computation, a quantum gate is a basic quantum circuit operating on
a small number of qubits. We show how to realize typical quantum gates on the qubit in the following. In some of the cases, the special state $|0\rangle$ can be used as a auxiliary state, or a bridge.

{\it $\frac{\pi}{4}$ phase shift gate}:
$|+\rangle+|-\rangle \longrightarrow |+\rangle +e^{i\frac{\pi}{2}}|-\rangle$.
In this case, we need only a phase factor $e^{i\frac{\pi}{2}}$ for the $|-\rangle$ term. This can be achieved by applying a unitary transformation $P_f(\frac{\pi}{2})$ on the starting state $|+\rangle+|-\rangle$. As a result, we obtain the evolution operator for the $\frac{\pi}{4}$ phase shift gate:
\begin{equation}
U_{\frac{\pi}{4}}=P_f(\frac{\pi}{2})e^{-i\pi}.
\end{equation}
The time duration $\tau_1$ and the time-independent magnetic field $h_1$ can be given by $\tau_1=\frac{\pi}{D} \sim 1.1$ns and $h_1=\frac{\sqrt{3}D}{2\gamma} \sim 888$G.

{\it $\frac{\pi}{8}$ phase shift gate}:
$|+\rangle+|-\rangle \longrightarrow |+\rangle +e^{i\frac{\pi}{4}}|-\rangle$.
It is similar to the $\frac{\pi}{4}$ phase shift gate. The phase factor is $e^{i\frac{\pi}{4}}$ in this case.
Applying $P_f(\frac{\pi}{4})$ on $(|+\rangle+|-\rangle)$, we obtain the evolution operator for the $\frac{\pi}{8}$ phase shift gate
\begin{equation}
U_{\frac{\pi}{8}}=P_f(\frac{\pi}{4})e^{-i\frac{\pi}{4}}.
\end{equation}
The time duration $\tau_2$ and the time-independent magnetic field $h_2$ are given by
$\tau_2=\frac{3\pi}{2D} \sim1.64$ns and $h_2=\frac{\sqrt{7}D}{6\gamma} \sim 452$G.

{\it Pauli-X gate}:
$|+\rangle\longrightarrow|-\rangle$.
This gate can be realized by applying $U_o^{\dag}(0,t_1)$ on the initial state $|+\rangle$ and then $U_o(\frac{\pi}{2},t_2)$ on the resulting intermediate state $|0\rangle$:
\begin{equation}
|-\rangle \propto U_o(\frac{\pi}{2},t_2) U_o^{\dag}(0,t_1) |+\rangle.
\end{equation}
In this way the gate is realized in two steps. Letting $\chi_1(t_1)=\chi_2(t_2)=\frac{\pi}{2}$, we have $t_1=t_2=T$, and then obtain $\kappa_1=\kappa_2=\frac{{15(\lambda T-\frac{\pi}{2})}}{{{\lambda^3}{T^3}}}$.
Using the phase condition in (\ref{eq15}), we can obtain $T=\frac{3.93}{D}$. For the first step, the time-dependent field function $J(t)/\lambda$ and the probability $P_{|+\rangle}$ of $|+\rangle$ as functions of time ($\lambda t$) are shown in Fig. 2. For the second step, we have similar time dependence for the field and the probability. After these two steps, the evolution operator for the Pauli-X gate is given by
\begin{equation}
V_X=U_o(\frac{\pi}{2},T) U_o^{\dag}(0,T)e^{i\pi/2}.
\end{equation}
And the total time interval is equivalent to $2T\sim 2.7$ns.

In addition, this state can be realized without applying $U_o(\theta_2,T_1)$ on state $|0\rangle$,
where $T_1$ is pulse time duration. At this time, we must guarantee matrix element $\bar{u}_{21}(T_1)$=$0$. After applying operator $U_o(\theta_2,T_1)$ on state $|+\rangle$, we derive
\begin{equation}
U_{o}(\theta_2,T_1)|+\rangle=d(T_1)\left(
\begin{array}{ccc}
e^{-i\frac{D}{2}T_1}\sin^2\theta_2-e^{-i\chi_3(T_1)}\cos^2\theta_2\\
0  \\
i(e^{-i\frac{D}{2}T_1}+e^{-i\chi_3(T_1)})\cos\theta_2\sin\theta_2 \\
\end{array}
\right)
\end{equation}
with $\xi_{\pm}(T_1)$=$\pi+2m\pi$ and $m$ is positive integer.
The condition to achieve state $|-\rangle$ is $\sin^2(2\theta_2)[1+\cos(\frac{D}{2}T_1-\chi_3(T_1))]$=$2$.
It can be satisfied by setting
$\cos(\frac{D}{2}T_1-\chi_3(T_1))$=$1$ and $\theta_2$=$\frac{\pi}{4}$. A reasonable result is
$\frac{D}{2}T_1-\chi_3(T_1)$=$2\pi$ and then $\kappa_3$=$\frac{30\pi}{\lambda^3T_1^3}$.
Using the phase condition about $\xi_{\pm}(T_1)$ in equation (\ref{eq15}) by replacing $\frac{\pi}{2}$ with $\pi+2m\pi$,
we can get a self-consistent value, $T_1\sim 5.3$ns from $\frac{D}{2}T_1 \approx7.542$ with $m=4$. Therefore, the evolution operator can be expressed as
\begin{equation}
U_X^{\prime}=U_o(\theta_2,T_1)e^{i(DT_1+\pi/2}.
\end{equation}
The smooth pulse and probability evolution of state $|+\rangle$ is shown in Fig. 3. Comparing it with Fig. 2, we can see that it is not as efficient as that using the intermediate state $|0\rangle$ because the magnetic field as a function of time is irregular.

\begin{figure}
\begin{center}
\includegraphics[width=0.8\columnwidth]{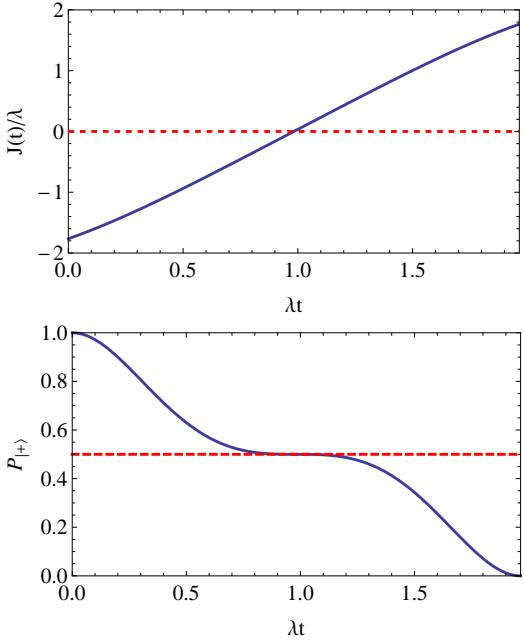}
\caption{\label{fig:qtest2} The time $t$ (in $1/\lambda$) dependence of $J(t)/\lambda$ (upper panel) and $P_{|+\rangle}$ (lower panel), with $\kappa=\frac{{15(\lambda
T_f-\frac{\pi}{2})}}{{{\lambda^3}{T_f^3}}}$ in Eq. (18).}
\end{center}
\end{figure}

\begin{figure}
\begin{center}
\includegraphics[width=0.8\columnwidth]{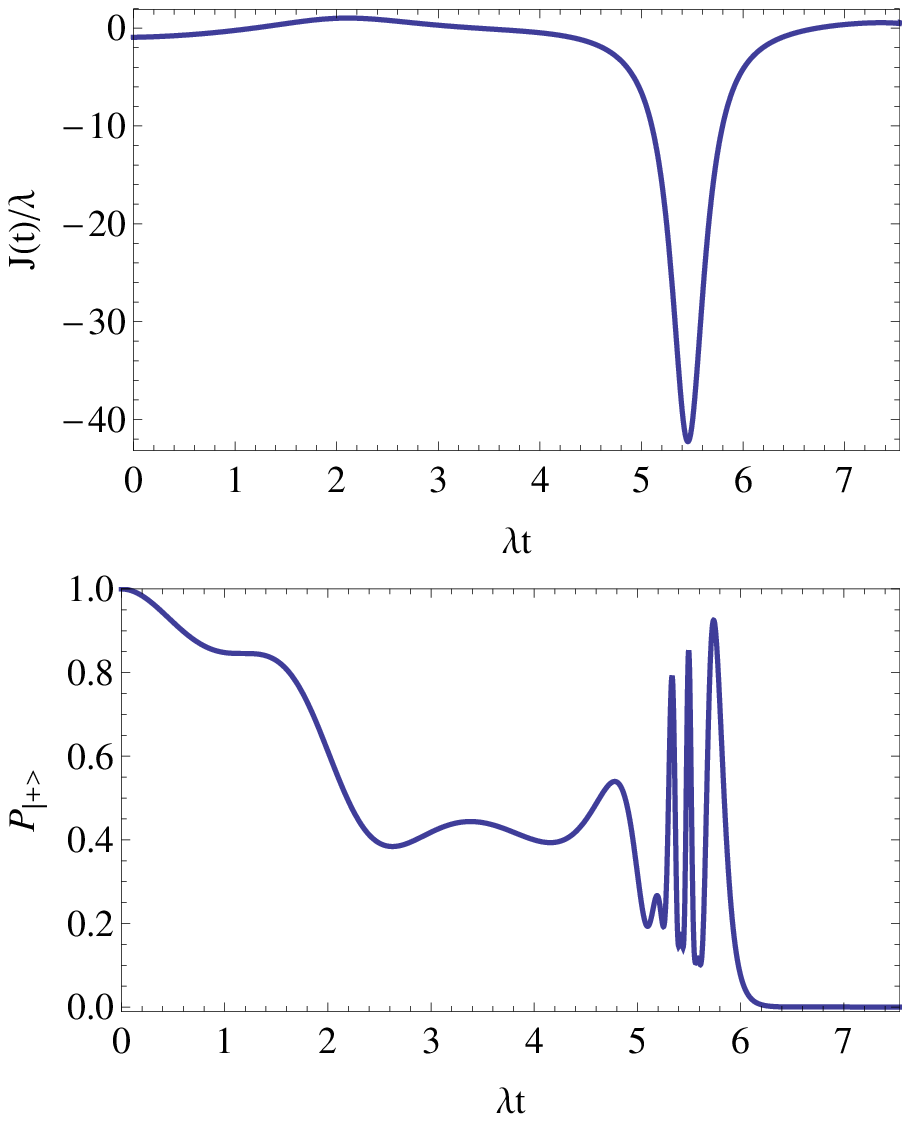}
\caption{\label{fig:qtest2} The time $t$ (in $1/\lambda$) dependence of $J(t)/\lambda$ (upper panel) and $P_{|+\rangle}$ (lower panel), with $\kappa=\frac{30\pi}{\lambda^3T_f^3}$ in Eq. (18).}
\end{center}
\end{figure}

{\it Hadamard gate}:
$|\pm\rangle\longrightarrow\frac{1}{\sqrt{2}}|+\rangle
\pm\frac{1}{\sqrt{2}}|-\rangle$.
In this situation, we can't use operator (\ref{ui}) directly. So in first step,
we change it to state $|0\rangle$. It is the same step mentioned in dealing with Pauli-X gate: $U_o^{\dag}(0,T) |+\rangle$.
After this step, for case $|0\rangle$ turning to $\frac{1}{\sqrt{2}}(|+\rangle+|-\rangle)$,
we only have to set $\theta_1=\frac{\pi}{4}$ and $\varphi=0$ in operator (\ref{ui}). The evolution operator for the Hadamard gate is
\begin{equation}
U_H=U_I(0,\frac{\pi}{4}) U_o^{\dag}(0,T).
\end{equation}
The total time is $2T+\tau_f\approx\frac{2\times3.93+\pi}{D} \approx 3.83$ns.

In addition, this gate can be realized without using $|0\rangle$. In this case, the static field is applied in $z$-axis, and the time evolution operator for the Hamiltonian (\ref{eq4})
with magnetic field $(0,0,B_z)$ is given by
\begin{equation}\label{eq30}
Q(t)=\left(
\begin{array}{ccc}
\cos(J_zt) & 0 & -i\sin(J_zt) \\
0 & e^{iDt} & 0 \\
-i\sin(J_zt) & 0 & \cos(J_zt)
\end{array}
\right)e^{-iDt}
\end{equation}
where $J_z$=$\gamma B_{z}$. If the initial state is $|+\rangle$ and $J_zt_f$=$\frac{\pi}{4}$, where $t_f$ is time duration,
$Q(t_f)$ makes the state $|+\rangle$ become $Q(t_f)|+\rangle=\frac{1}{\sqrt{2}}(|+\rangle-i|-\rangle)$.
In order to get target state $\frac{1}{\sqrt{2}}(|+\rangle+|-\rangle)$, we need a phase factor $e^{i\pi/2}$ for the $|-\rangle$.
It can be shown that the time value and magnetic field is the same as
$\tau_1$ and $h_1$. With the phase shift operator $P_f(\frac{\pi}{2})$,
the whole process can be represented as
\begin{equation}\label{eq32}
U_H^{\prime}
=P_f(\frac{\pi}{2})Q(t_f)=\left(\begin{array}{ccc}
i\frac{1}{\sqrt{2}} & 0 & \frac{1}{\sqrt{2}} \\
0 & ie^{iDt_f} & 0 \\
i\frac{1}{\sqrt{2}} & 0 & -\frac{1}{\sqrt{2}}%
\end{array}
\right)e^{-iDt_f}.
\end{equation}
The total time duration is $t_f+\tau_1=
\frac{\pi}{4\gamma B_{z}}+\frac{\pi}{D} \approx 1.375$ns with $B_z \sim 1000$G. Furthermore, this gate can change $|-\rangle$ into $\frac{1}{\sqrt{2}}|+\rangle-\frac{1}{\sqrt{2}}|-\rangle$.

{\it Arbitrary gating}. In this case, we need to realize
$\cos\theta_{1}|+\rangle +e^{i\varphi _{1}}\sin \theta
_{1}|-\rangle \longrightarrow \cos \theta_{2}|+\rangle+e^{i\varphi
_{2}}\sin \theta_{2}|-\rangle$.
Assuming $0\leq\theta_{1},\theta_{2}\leq\pi$ and
$0\leq\varphi_{1},\varphi_{2}\leq \pi$. If
$\theta_{1}$=$\theta_{2}$, we need only to modify the relative phase.
According to the previous subsection, we can realize this control
through the intermediate state $|0\rangle$. The first part is actually the
inverse process of that in Sec. \textrm{IV}. As the operator (\ref{ui})
is unitary, the first process can be achieved by the operator $U_I^{\dag}(\varphi_1,\theta_1)$,
with $T_2$ being the pulse time duration. Then, the target state can be realized by
$U_I(\varphi_2,\theta_2)$ with $T_3$ being the pulse time duration. The whole process can be represented by
\begin{equation}\label{eq33}
U_A(\varphi_2, \theta_2,\varphi_1,\theta_1)
=U_I(\varphi_2,\theta_2)U_I^{\dag}(\varphi_1,\theta_1)
\end{equation}

This gate can be realized without the intermediate state $|0\rangle$. It is clear from operator (\ref{eq30}) that  $J_zt$ can be seen as one variable because the overall phase factor does not play role in the gate. As a result, the larger  the longitudinal field, the smaller the time duration. The initial relative phase $\varphi_1$ should be adjusted before the probability amplitude is changed.
At first, we change the relative phase $\varphi_{1}$ into $\frac{\pi}{2}$ by $P_f^{\dag}(\varphi_1-\frac{\pi}{2})$, making a state $|\Phi(0)\rangle=\cos\theta_1|+\rangle
+i\sin\theta_1|-\rangle$. Then, we apply the time evolution operator $Q(t_e)$ on $|\Phi(0)\rangle$, where $t_e$ is the time duration, and obtain the state
$|\Phi(t_e)\rangle=(\cos(\theta_1-J_z t_e)|+\rangle+i\sin(\theta_1-J_z t_e)|-\rangle)e^{-iDt_e}$.
Here, $J_z t_e$ can be expressed as
\begin{equation}
J_zt_e=\left\{\begin{array}{ll}\theta_1-\theta_2 & ~{\bf if}~~ \theta_1\ge \theta_2\\ \theta_1-\theta_2+2\pi & ~{\bf if}~~ \theta_1< \theta_2\end{array}\right.
\end{equation}
Finally, we apply the phase regulation $P_f(\varphi_2-\frac{\pi}{2})$ to get the phase $\varphi_2$.
The whole procedure can be represented as the unitary operator:
\begin{equation}\label{eq34}
U^{\prime}_A(\varphi_2, \theta_2,\varphi_1,\theta_1) =P_f(\varphi_2-\frac{\pi}{2})Q(t_e)P_f^{\dag}(\varphi_1-\frac{\pi}{2})
\end{equation}

\begin{table}[h]
\caption{The time durations ($T_G$) of the $\frac{\pi}{4}$ phase, $\frac{\pi}{8}$ phase, Pauli-X, and Hadamard gates}
\begin{ruledtabular}
\begin{tabular}{c|cccc}
Gate & $\frac{\pi}{4}$ phase & $\frac{\pi}{8}$ phase & Pauli-X & Hadamard \\
\hline
$T_G$ & 1.1 ns   &  1.6 ns   &  2.7 ns   & 3.8 ns \\
      &         &           &  5.3 ns   &  1.4 ns
\end{tabular}
\end{ruledtabular}
\end{table}

\section{Discussion and Conclusion}

The time durations $T_G$ of the typical gates are summarized in Table I. It is clear that the gates need at most a few nanoseconds. Furthermore, it can be estimated to take only nanoseconds to complete the initialization and gating of arbitrary qubits. Therefore, the qubits on the basis of $|+\rangle$ and $|-\rangle$ can be fast initialized and gated with magnetic fields, in comparison to those in terms of $|+1\rangle$ and $|-1\rangle$ with strong mechanical driving\cite{mech}. When the magnetic field is applied along the z axis, the effect of the time evolution
operator (\ref{eq30}) on our NV center qubit with the basis of $|+\rangle$ and $|-\rangle$ looks like that on spin-$\frac 12$ qubit in ``bang-bang'' approach\cite{zhihui}, but they are different from each other. Because here one needs to control the magnetic field strength only, the theoretical fidelity can be very high for these gates.

In summary, by choosing $|\pm \rangle$, defined as $(|+1\rangle \pm|-1\rangle)/\sqrt{2}$, of the NV center spin as the qubit basis, we have obtained exact evolution operator of the NV center spin under time-dependent magnetic field by mapping the three-level system of the NV center spin on a two-level spin system under a time-dependent magnetic field and using the existing exact analytical results of the quantum two-level system\cite{barnes,barnes1}. With the exact evolution operator of the NV center spin including three levels, we have shown how to pre pare arbitrary qubits with the basis $|\pm \rangle$  from the starting $|0\rangle$ state and to realize arbitrary rapid quantum logic gates for the qubits. It also has been estimated that the typical quantum logic gates can be accomplished within a few nanoseconds. We believe that the fidelity can be very high because only magnetic field strength needs to be controlled. these results are useful to realizing quantum computing with the NV center spin systems in diamond and exploring other quantum effects and applications.

\begin{acknowledgments}
This work is supported by Nature Science Foundation of China (Grant Nos. 11174359 and 11574366), by Chinese Department of Science and Technology
(Grant No. 2012CB932302), and by the Strategic Priority Research
Program of the Chinese Academy of Sciences (Grant No. XDB07000000).
\end{acknowledgments}

\end{document}